\documentclass{ws-mpla}

\begin{document}

\markboth{K. Konno, T. Matsuyama \& S. Tanda}
{EFFECTS OF CS CORRECTIONS
ON CONSERVED QUANTITIES OF FLUID}

\catchline{}{}{}{}{}

\title{EFFECTS OF CHERN-SIMONS CORRECTIONS
ON CONSERVED QUANTITIES OF RELATIVISTIC FLUID
}

\author{KOHKICHI KONNO}

\address{Department of Natural and Physical Sciences, 
   Tomakomai National College of Technology \\ 
   Tomakomai 059-1257, Japan \\
   kohkichi@gt.tomakomai-ct.ac.jp}

\author{TOYOKI MATSUYAMA}
\address{Department of Physics, Nara University of Education\\
         Nara 630-8528, Japan
}

\author{SATOSHI TANDA}
\address{Center of Education and Research for Topological 
   Science and Technology\\ Department of Applied Physics, 
   Hokkaido University\\ Sapporo 060-8628, Japan}

\maketitle

\pub{Received (Day Month Year)}{Revised (Day Month Year)}

\begin{abstract}
 We consider relativistic fluid flow under Chern-Simons
 modified Maxwell theory and under Chern-Simons modified 
 gravity theory. We take account of the effects of Chern-Simons 
 corrections on the quantities of fluid flow that are conserved 
 without the Chern-Simons corrections. We find that the conservations 
 of several quantities are generally broken by 
 the Chern-Simons corrections.

\end{abstract}

\ccode{PACS Nos.: include PACS Nos.}

\section{Introduction}	

The study of symmetry is requisite in modern physics. 
A typical example can be found in the standard theory for 
the interactions between elementary particles.
In contrast, the discoveries of symmetry breaking phenomena 
have opened new frontiers in physics.\cite{qsb1,qsb2,sc1,sc2}
Further detections of such phenomena
might provide a signature of new physics.
Chern-Simons (CS) modified theories\cite{cfj,jp} are candidates 
for the theories that describe the breaking of parity symmetry.
The CS modified theories have attracted a lot of attentions 
in the context of exploring new physics.\cite{lwk,cpt1,cpt2,cpt3}
The CS modified Maxwell theory\cite{cfj} is constructed from 
the usual electromagnetic action with a CS term.
Similarly, the CS modified gravity theory\cite{jp}
is constructed from the Einstein-Hilbert action
with a CS term. The CS terms violate the parity symmetry.
In previous works, electromagnetic and 
gravitational fields have mainly been 
investigated.\cite{cf1,cf2,cf3,cf4,ay1,ay2,kmt1,seck,kmat,kmt2,cs-review}
In this paper, we discuss the effect of the 
symmetry breaking terms on fluid flow 
under the CS modified theories.

This paper focuses on conserved quantities of fluid. 
For its discussion, it is useful to adopt 
the Lagrangian description,\cite{asada-kasai,asada} 
in which each fluid particle is assigned a label, 
i.e., Lagrangian coordinates at the initial time. 
In the Lagrangian coordinates, mass evaluated in a closed 
region is conserved under time evolution. Similarly, 
for perfect fluid, circulation is also conserved. 
These facts are valid even when we take account 
of general relativity.\cite{asada-kasai,asada} 
Furthermore, we can encounter other conserved quantities, 
e.g., fluid helicity,\cite{helicity1,helicity2,fukumoto,bekenstein} 
magnetic helicity and 
cross helicity.\cite{woltjer1,woltjer2,woltjer3,bekenstein}
The last two quantities are considered 
for fluid interacting with electromagnetic fields.
In this paper, we deal with the above-mentioned conserved 
quantities within a general relativistic framework
and take account of the corrections induced by the 
CS modified theories.

This paper is organized as follows.
In Sec.~\ref{sec:cs-maxwell}, we consider fluid flow interacting 
with electromagnetic fields under the CS modified Maxwell theory.
Utilizing the Lagrangian description, we discuss the conservation 
of mass (or energy), circulation, and fluid helicity.
We also investigate magnetic helicity and cross helicity. 
In Sec.~\ref{sec:cs-gravity}, we deal with fluid flow 
under the CS modified gravity theory. 
We discuss the conserved quantities in the same way 
as in Sec.~\ref{sec:cs-maxwell}.
Finally we provide a summary in Sec.~\ref{sec:summary}. 
Throughout the paper, we use geometrized units with $c=G=1$.

\section{Fluid Flow in CS Modified Maxwell Theory}
\label{sec:cs-maxwell}

\subsection{CS modified Maxwell theory}

The action of the CS modified Maxwell theory 
is provided by\cite{cfj}
\begin{eqnarray}
\label{eq:cs-m-action}
 I_{\rm EM} & = & \int d^4 x 
  \left( - \frac{1}{16\pi} F_{\mu\nu} F^{\mu\nu} 
   + \frac{1}{8\pi} v_{\mu} 
   \ ^{\ast} F^{\mu\nu} A_{\nu} \right) , 
\end{eqnarray}
where $A_{\mu}$ is the four-potential, 
$F_{\mu\nu} \equiv \partial_{\mu} A_{\nu} - \partial_{\nu} A_{\mu}$
is the field-strength tensor, $^{\ast} F^{\mu\nu} \equiv
\frac{1}{2} \varepsilon^{\mu\nu\lambda\sigma} F_{\lambda\sigma}$
is the dual field-strength tensor, and $v_{\mu}$ is an external 
four-vector called the embedding vector. Here, 
$\varepsilon^{\mu\nu\lambda\sigma}$ denotes the Levi-Civita tensor
with $\varepsilon^{0123}=1$. Raising and lowering 
the indices of tensors are done by the Minkowski metric 
$\eta_{\mu\nu}$ and $\eta^{\mu\nu}$.
The second term in the integrand in Eq.~(\ref{eq:cs-m-action})
is called the CS term. 
The embedding vector $v_{\mu}$ is now assumed to satisfy 
$\partial_{\mu} v_{\nu} = 0$ to ensure gauge invariance.\cite{cfj} 
The dual field-strength tensor satisfies the Bianchi identity 
$\partial_{\mu} \: ^{\ast} F^{\mu\nu} = 0$.
The variation of the action with respect to $A_{\mu}$ gives
the left-hand side of the electromagnetic field equation
\begin{equation}
\label{eq:maxwell}
 \partial_{\nu} F^{\mu\nu} + v_{\nu}
 \: ^{\ast} F^{\mu\nu} = 4\pi J^{\mu} ,
\end{equation}
where $J^{\mu}$ is the electric four-current. The second term on the 
left-hand side in this equation stems from the CS term 
in Eq.~(\ref{eq:cs-m-action}). Thus the field equation 
of the Maxwell theory is modified by the CS correction 
as in Eq.~(\ref{eq:maxwell}).

\subsection{Basic equations for fluid flow}

The basic equations for fluid flow under the CS modified 
Maxwell theory can be obtained from the 
conservation law of energy-momentum tensor $T^{\mu\nu}$.
The energy-momentum tensor is composed of the fluid part 
$T_{\rm m}^{\ \mu\nu}$ and the electromagnetic part 
$T_{\rm em}^{\quad \mu\nu}$, i.e.,
\begin{equation}
\label{eq:mhd-e-m}
 T^{\mu\nu} = T_{\rm m}^{\ \mu\nu} 
  + T_{\rm em}^{\quad \mu\nu} .
\end{equation}
For the fluid part, we assume perfect fluid 
\begin{equation}
 T_{\rm m}^{\ \mu\nu} = \left( \rho + p \right) u^{\mu} u^{\nu}
 + p \eta^{\mu\nu} , 
\end{equation}
where $\rho$ is the energy density, $p$ is the pressure, and 
$u^{\mu}$ denotes the four-velocity field of fluid particles
($u^{\mu}u_{\mu} =-1$).
In the CS modified Maxwell theory, the electromagnetic part 
$T_{\rm em}^{\quad \mu\nu}$ is given by\cite{cfj}
\begin{equation}
\label{eq:em-em}
 T_{\rm em}^{\quad \mu\nu} = \frac{1}{4\pi}
 \left( F^{\mu\lambda} F^{\nu}_{\ \ \lambda} 
 - \frac{1}{4} \eta^{\mu\nu} F_{\lambda\sigma} F^{\lambda\sigma}
 + \frac{1}{2} v^{\nu} \: ^{\ast} F^{\mu\lambda} A_{\lambda} \right) .
\end{equation}
 From the conservation equation $\partial_{\mu} T^{\mu\nu} = 0$,
using Eqs.~(\ref{eq:maxwell}), (\ref{eq:mhd-e-m}) 
and (\ref{eq:em-em}), we obtain
\begin{equation}
 \partial_{\mu} T_{\rm m}^{\ \mu\nu} 
 = F^{\nu\lambda} J_{\lambda} .
\end{equation}
This equation can be divided into two parts, i.e.,
the component parallel to $u^{\mu}$ and 
the components orthogonal to $u^{\mu}$.
The former gives the continuity equation
\begin{equation}
\label{eq:mhd-c-e1}
 \left( \rho u^{\mu} \right)_{, \mu}
 + p u^{\mu}_{\ ,\mu} = 0 ,
\end{equation}
where a comma denotes the partial differentiation 
with respect to coordinates.
The latter gives the equation of motion 
\begin{equation}
\label{eq:mhd-eom1}
 \left( \rho + p \right) u^{\nu} u^{\mu}_{\ , \nu}
 + P^{\mu\nu} p_{,\nu} = F^{\mu\nu} J_{\nu} ,
\end{equation}
where $P^{\mu\nu} \equiv \eta^{\mu\nu} + u^{\mu} u^{\nu}$ 
is the projection tensor. In deriving Eqs.~(\ref{eq:mhd-c-e1})
and (\ref{eq:mhd-eom1}), we used the ideal magnetohydrodynamics
approximation $u_{\mu} F^{\mu\nu} = 0$. 
We now assume barotropic fluid, for which 
new variables $s$ and $h$ can be introduced as\cite{ehlers,asada}
\begin{equation}
\label{eq:entropy}
 s \equiv \exp \left( \int^{\rho} \frac{d\rho}{\rho +p} \right) ,
\end{equation}
\begin{equation}
\label{eq:enthalpy}
 h \equiv \exp \left( \int^{p} \frac{dp}{\rho +p} \right) .
\end{equation}
The functions $s$ and $h$ correspond to the specific entropy
and the specific enthalpy, respectively.\cite{ehlers,asada}
When the pressure vanishes, we have $s=\rho$ and $h=1$ 
by taking appropriate constants of integration in the integrals
in Eqs.~(\ref{eq:entropy}) and (\ref{eq:enthalpy}).
By using $s$ and $h$, Eqs.~(\ref{eq:mhd-c-e1}) and 
(\ref{eq:mhd-eom1}) are rewritten, respectively, as
\begin{equation}
\label{eq:mhd-c-e2}
  \left( s u^{\mu} \right)_{, \mu} = 0 ,
\end{equation}
\begin{equation}
\label{eq:mhd-eom2}
 u^{\nu} u^{\mu}_{\ , \nu}
 + P^{\mu\nu} \left( \ln h \right)_{,\nu} 
 = N^{\mu} - V^{\mu} , 
\end{equation}
where 
\begin{eqnarray}
 N^{\mu} & \equiv & \frac{1}{4\pi \left( \rho + p \right)}
   F^{\mu}_{\ \ \lambda} F^{\lambda\sigma}_{\quad , \sigma} , \\
 V^{\mu} & \equiv & \frac{1}{16\pi \left( \rho + p \right)}
   \: ^{\ast} F^{\lambda\sigma} F_{\lambda\sigma} v^{\mu} . 
\end{eqnarray}
To derive Eq.~(\ref{eq:mhd-eom2}), we used the identity
\begin{equation}
\label{eq:identity-F}
 ^{\ast} F^{\mu\lambda} F_{\nu\lambda} = \frac{1}{4} 
   \delta^{\mu}_{\ \nu} F^{\lambda\sigma} F_{\lambda\sigma} ,
\end{equation}
where $\delta^{\mu}_{\ \nu}$ denotes the Kronecker delta.
In Eq.~(\ref{eq:mhd-eom2}), $V^{\mu}$ comes from the CS correction. 
Thus while the continuity equation (\ref{eq:mhd-c-e2}) is unchanged, 
the equation of motion (\ref{eq:mhd-eom2}) is changed due to the 
CS correction.

We also discuss vorticity of fluid. 
For this purpose, let us define the four-vorticity 
$\omega^{\mu}$ as 
\begin{equation}
\label{eq:4-v}
 \omega^{\mu} \equiv \frac{1}{2} \varepsilon^{\mu\nu\lambda\sigma}
   u_{\nu} u_{\lambda , \sigma} .
\end{equation}
The spatial components of the four-vorticity give the usual 
three-vorticity $\nabla \times \mbox{\boldmath $v$}_{\rm f}$ 
when the motion is non-relativistic, 
i.e., $u^{\mu} \simeq \left( 1, v_{\rm f}^{\ i}\right)$ and 
$\left| \mbox{\boldmath $v$}_{\rm f} \right| \ll 1$, 
where the Latin index $i$ runs over the spatial coordinates.  
We emphasize that the temporal component of $\omega^{\mu}$
gives the density of fluid helicity  
$\mbox{\boldmath $v$}_{\rm f} \cdot \left( \nabla \times 
\mbox{\boldmath $v$}_{\rm f} \right)$ in the non-relativistic 
case (see also Refs.~\refcite{helicity1,helicity2,bekenstein}). 
Hence, $\omega^{\mu}$ may be 
regarded as the four-current of fluid helicity.
This fact gives us a new insight that we can treat both
helicity and circulation of fluid in a unified way
by adopting the four-vorticity $\omega^{\mu}$
defined in Eq.~(\ref{eq:4-v}).
Furthermore, we can easily recognize the transformation law
of helicity density under a coordinate transformation.
Under a coordinate transformation $x^{\mu} \rightarrow x'^{\mu}$, 
the helicity density $\omega^{0}$ is transformed 
according to the transformation law
$\omega^{\mu} \rightarrow \omega^{\mu'} = 
\left( \partial x'^{\mu} /\partial x^{\nu} \right) \omega^{\nu}$.
 From Eq.~(\ref{eq:mhd-eom2}), we obtain
the differential equation for $\omega^{\mu}$, 
\begin{eqnarray}
\label{eq:mhd-v-e}
 \left( \frac{h\omega^{\mu}}{s} \right)_{,\nu}
  \frac{u^{\nu}}{h} - \frac{h\omega^{\nu}}{s} 
  \left( \frac{u^{\mu}}{h}\right)_{,\nu}
 = \frac{1}{s} \left[ \left( N_{\nu}  \omega^{\nu} u^{\mu} 
  + \: ^{\ast} M^{\mu\nu} u_{\nu} \right) 
  - \left( V_{\nu} \omega^{\nu} u^{\mu} 
  + \: ^{\ast} W^{\mu\nu} u_{\nu} \right) \right] , 
 \nonumber \\
\end{eqnarray}
where 
\begin{eqnarray}
 ^{\ast} M^{\mu\nu} & \equiv & \frac{1}{2} 
   \varepsilon^{\mu\nu\lambda\sigma} N_{\lambda , \sigma} , \\
 ^{\ast} W^{\mu\nu} & \equiv & \frac{1}{2} 
   \varepsilon^{\mu\nu\lambda\sigma} V_{\lambda , \sigma} .
\end{eqnarray}
Here $^{\ast} M^{\mu\nu}$ and $^{\ast} W^{\mu\nu}$ are interpreted 
as rotational parts of the derivatives of
$N_{\mu}$ and $V_{\mu}$, respectively.
Thus when an electromagnetic field or the CS correction exists, 
the vorticity equation becomes an inhomogeneous equation 
as seen in Eq.~(\ref{eq:mhd-v-e}).

Consequently, the basic equations for fluid flow under CS modified Maxwell 
theory are given by the field equation (\ref{eq:maxwell}), 
the continuity equation (\ref{eq:mhd-c-e2}), the equation of motion 
(\ref{eq:mhd-eom2}) and the vorticity equation (\ref{eq:mhd-v-e}).

\subsection{Lagrangian description of fluid flow}

We deal with the fluid motion from the viewpoint of 
the Lagrangian description.
We adopt the Lagrangian coordinates\cite{asada-kasai,asada}
\begin{equation}
 x^{\mu} = \left( \tau , x^{i} \right) 
  = \left( \tau , \mbox{\boldmath $x$} \right),
\end{equation}
where $\tau$ is the proper time of a fluid particle
and $x^{i}$ is constant along a line of fluid flow. 
In this coordinates, the four-velocity becomes
\begin{equation}
\label{eq:L-c}
 u^{\mu} = \frac{dx^{\mu}}{d\tau}
  = \delta^{0\mu} = \left( 1 , 0 , 0, 0 \right) .
\end{equation} 
Adopting the Lagrangian coordinates is equivalent 
to taking the four-velocity in the form of Eq.~(\ref{eq:L-c}).
Equation (\ref{eq:L-c}) is called as 
the Lagrangian condition.\cite{asada-kasai,asada}
When we use the Lagrangian condition, 
the metric $\eta_{\mu\nu}$ and the partial derivative $\partial_{\mu}$ 
in Eqs.~(\ref{eq:maxwell}), (\ref{eq:mhd-c-e2}), 
(\ref{eq:mhd-eom2}) and (\ref{eq:mhd-v-e})
must be replaced with the general metric $g_{\mu\nu}$ and 
the covariant derivative $\nabla_{\mu}$, respectively,
because the coordinates are no longer Cartesian.
Furthermore, the Levi-Civita tensor is redefined as
$\varepsilon^{0123} = 1/\sqrt{-g}$, 
where $g$ is the determinant of $g_{\mu\nu}$. 
In the Lagrangian description, 
the metric is changed with fluid motion,
while the coordinates of fluid particles are fixed.

Let us discuss the continuity equation (\ref{eq:mhd-c-e2})
using the Lagrangian condition (\ref{eq:L-c}). 
Equation (\ref{eq:mhd-c-e2}) gives 
\begin{equation}
 \left( \sqrt{-g} s\right)_{,0} = 0 .
\end{equation}
Thus we obtain
\begin{equation}
\label{eq:mhd-s}
 s(\tau , \bm{x}) = \sqrt{\frac{g(\tau_{0},\bm{x})}{g(\tau,\bm{x})}}
  s(\tau_{0} , \bm{x}) . 
\end{equation}
When the pressure is negligible, Eq.~(\ref{eq:mhd-s}) reduces to
\begin{equation}
\label{eq:mhd-rho} 
 \rho(\tau , \bm{x}) = \sqrt{\frac{g(\tau_{0},\bm{x})}{g(\tau,\bm{x})}}
  \rho(\tau_{0} , \bm{x}) ,  
\end{equation}
Therefore, Eqs.~(\ref{eq:mhd-s}) and (\ref{eq:mhd-rho}) 
lead to the conservations of entropy density $\sqrt{-g} s$ 
and energy density $\sqrt{-g} \rho$, respectively, 
in the Lagrangian coordinates.

Next we discuss the vorticity equation (\ref{eq:mhd-v-e}) 
using the Lagrangian condition (\ref{eq:L-c}). 
Here it should be noted that the temporal component of $\omega^{\mu}$
is not independent of the spatial components
because we have $\omega^{0} = g_{0i}\omega^{i}$
from the equality $u_{\mu} \omega^{\mu} = 0$.
Hence we focus on the spatial components of $\omega^{\mu}$.
The spatial components of Eq.~(\ref{eq:mhd-v-e}) give
\begin{equation}
 \left( \frac{h\omega^{i}}{s} \right)_{,0}
 = \frac{h}{s} \left( \!\: ^{\ast} M^{i\nu} 
  - \: ^{\ast} W^{i\nu} \right) u_{\nu} .
\end{equation}
Thus we obtain
\begin{eqnarray}
\label{eq:mhd-h-omega}
 \lefteqn{h(\tau , \bm{x}) \omega^{i}(\tau , \bm{x})} \nonumber \\
 & = & \sqrt{\frac{g(\tau_{0},\bm{x})}{g(\tau,\bm{x})}}
  \left[ 
  h(\tau_{0} , \bm{x}) \omega^{i}(\tau_{0} , \bm{x}) 
  + s(\tau_{0} , \bm{x}) \int_{\tau_{0}}^{\tau} d\tau' 
   \frac{h}{s} \left( \!\: ^{\ast} M^{i\nu} 
  - \: ^{\ast} W^{i\nu} \right) u_{\nu} \right] ,
\end{eqnarray}
where we have used Eq.~(\ref{eq:mhd-s}).
When the pressure is negligible, 
Eq.~(\ref{eq:mhd-h-omega}) reduces to
\begin{eqnarray}
\label{eq:mhd-omega}
 \omega^{i}(\tau , \bm{x})
 & = & \sqrt{\frac{g(\tau_{0},\bm{x})}{g(\tau,\bm{x})}}
  \left[ 
  \omega^{i}(\tau_{0} , \bm{x}) + \rho (\tau_{0} , \bm{x})
  \int_{\tau_{0}}^{\tau} d\tau' 
  \frac{1}{\rho} \left( \: ^{\ast} M^{i\nu} 
  - \: ^{\ast} W^{i\nu} \right) u_{\nu} \right] .
\end{eqnarray}
Equation (\ref{eq:mhd-h-omega}) (or (\ref{eq:mhd-omega}))
means that the spatial components 
$\sqrt{-g} h\omega^{i}$ (or $\sqrt{-g} \omega^{i}$)
is conserved in the absence of both the rotational part 
of electromagnetic force and the CS correction. 
In particular, when the pressure is negligible, 
we can obtain the conservation of
four-vector density $\sqrt{-g} \omega^{\mu}$ 
directly from Eq.~(\ref{eq:mhd-v-e}). 
In such a case, therefore, the conservations of helicity and 
circulation can be obtained in a unified way 
within a relativistic framework.
In the presence of the rotational part 
of electromagnetic force or the CS correction,
the conservation law concerning vorticity is broken. 
The integral terms in Eqs.~(\ref{eq:mhd-h-omega})
and (\ref{eq:mhd-omega}), which are independent 
of $\omega^{i}$, become sources of three-vorticity. 
It means that even if $\omega^{i}$ vanishes everywhere 
at the initial time, the spatial components of vorticity 
may be created at some time. 
The temporal component of the four-vorticity can be derived 
from $\omega^{0} = g_{0i} \omega^{i}$ as mentioned above.
When the pressure vanishes, $\omega^{0}$ 
is obtained directly from Eq.~(\ref{eq:mhd-v-e}),
\begin{eqnarray}
\label{eq:mhd-omega0}
 \omega^{0}  (\tau ,\bm{x}) 
 & = & \sqrt{\frac{g (\tau_{0} ,\bm{x})}{g (\tau ,\bm{x})}} 
   \left[ \omega^{0}(\tau_{0} , \bm{x})
   + \rho (\tau_{0} ,\bm{x}) \int_{\tau_{0}}^{\tau} d\tau'
   \frac{1}{\rho} 
   \right. \nonumber \\
 && \times \left.
    \left\{ \left( N_{\nu} \omega^{\nu}
    + \: ^{\ast} M^{0i} u_{i} \right)
    - \left( V_{\nu} \omega^{\nu}
    + \: ^{\ast} W^{0i} u_{i} \right) \right\} \right] .
\end{eqnarray}
The terms $^{\ast} M^{0i} u_{i}$ and $^{\ast} W^{0i} u_{i}$
in the integrand are independent of $\omega^{\mu}$.
This fact means that helicity density may also be created 
irrespective of the values of $\omega^{\mu}$.
Therefore, not only the conservation of circulation but also the 
conservation of helicity is generally 
broken in the CS modified Maxwell theory.

\subsection{Magnetic helicity and cross helicity}

Let us discuss magnetic helicity and cross helicity 
in the CS modified Maxwell theory.
These quantities are conserved under certain conditions
in ordinary magnetohydrodynamics.
In this subsection, we use the Cartesian coordinates, 
in which the metric reduces to the flat Minkowski metric.

We define the magnetic helicity four-current 
$H^{\ \ \mu}_{\rm m}$ as
\begin{equation}
 H^{\ \ \mu}_{\rm m} 
  \equiv \: ^{\ast} F^{\mu\nu} A_{\nu} .
\end{equation}
Using the identity (\ref{eq:identity-F}), we derive
for ideal magnetohydrodynamics
\begin{equation}
 ^{\ast} F^{\mu\nu} F_{\mu\nu}
 = - 4 u_{\mu} \: ^{\ast} F^{\mu\lambda} u^{\nu} F_{\nu\lambda}
 = 0 .
\end{equation}
Thus we obtain
\begin{equation}
 H^{\ \ \mu}_{{\rm m} \ \;  ,\mu} 
 = \frac{1}{2} \: ^{\ast} F^{\mu\nu} F_{\mu\nu} = 0 .
\end{equation}
Therefore, the magnetic helicity four-current $H^{\ \ \mu}_{\rm m}$ 
is conserved even in the CS modified Maxwell theory.

Next we discuss the cross helicity four-current 
$H^{\ \mu}_{\rm c}$ defined by
\begin{equation}
  H^{\ \mu}_{\rm c} 
  \equiv \: ^{\ast} \omega^{\mu\nu} A_{\nu} , 
\end{equation}
where $^{\ast} \omega^{\mu\nu} \equiv \frac{1}{2} 
\varepsilon^{\mu\nu\lambda\sigma} u_{\lambda , \sigma}$.
For ideal magnetohydrodynamics, we obtain
\begin{equation}
\label{eq:c-h}
 H^{\ \mu}_{{\rm c} \ \; ,\mu} 
  = \frac{1}{2} \: ^{\ast} \omega^{\mu\nu} F_{\mu\nu}
  = \frac{1}{2} B^{\mu} \left[ \left( \ln h \right)_{,\mu}
    + V_{\mu}\right] ,
\end{equation}
where $B^{\mu} \equiv \: ^{\ast} F^{\mu\nu} u_{\nu}$ 
denotes the magnetic field. The last term of the right-hand side,
which is proportional to $B^{\mu}V_{\mu}$, 
arises from the CS correction. When the right-hand side 
in Eq.~(\ref{eq:c-h}) does not vanish, 
the conservation of cross helicity is broken.
Therefore, the cross helicity four-current is not generally conserved 
in the CS modified Maxwell theory.

\section{Fluid Flow in CS Modified Gravity}
\label{sec:cs-gravity}

\subsection{CS modified gravity}

The action of the dynamical CS modified gravity 
is provided by\cite{jp,seck} 
\begin{eqnarray}
\label{eq:lagrangian}
 I_{\rm G} & = & \int d^4 x \sqrt{-g}
  \left[ - \frac{R}{16\pi} + \frac{\ell}{64\pi} 
  \vartheta \: ^{\ast}\!R^{\tau \ \mu\nu}_{\ \sigma}
  R^{\sigma}_{\ \tau\mu\nu}
  - \frac{1}{2} g^{\mu\nu} \left( \partial_{\mu} \vartheta \right)
  \left( \partial_{\nu} \vartheta \right) 
  - V\left( \vartheta \right)
  + {\cal L}_{\rm m} \right] , \nonumber \\
 &&
\end{eqnarray}
where $R \equiv g^{\alpha\beta} R_{\alpha\beta}$ 
($R_{\alpha\beta} \equiv R^{\lambda}_{\ \alpha\lambda\beta}$)
is the Ricci scalar,
$R^{\tau}_{\ \sigma\alpha\beta} \equiv \partial_{\beta}
\Gamma^{\tau}_{\sigma\alpha} - \cdots$ is the Riemann tensor
($\Gamma^{\alpha}_{\beta\gamma}$ is the Christoffel symbols),
$\ell$ is a coupling constant, $\vartheta$ is a dynamical scalar field,
$V$ is a potential, and ${\cal L}_{\rm m}$ is 
the Lagrangian density for matter.
The dual Riemann tensor is defined by
$^{\ast} \!R^{\tau\ \mu\nu}_{\ \sigma} \equiv \frac{1}{2}
\varepsilon^{\mu\nu\alpha\beta} R^{\tau}_{\ \sigma\alpha\beta}$,
where $\varepsilon^{0123} \equiv 1/ \sqrt{-g}$.
The second term in Eq.~(\ref{eq:lagrangian}) is 
called a Chern-Pontryagin term and
connected to the CS term, via partial integration,
\begin{equation}
 I_{\rm CS} = - \frac{\ell}{32\pi} \int d^{4}x
  \sqrt{-g} \left( \partial_{\mu} \vartheta \right)
  \varepsilon^{\mu\alpha\beta\gamma}
  \left( \Gamma^{\sigma}_{\alpha\tau} 
  \partial_{\beta} \Gamma^{\tau}_{\gamma\sigma}
  + \frac{2}{3}  \Gamma^{\sigma}_{\alpha\tau}
  \Gamma^{\tau}_{\beta\eta} \Gamma^{\eta}_{\gamma\sigma} \right).
\end{equation}
Thus nontrivial $\partial_{\mu} \vartheta $
leads to the CS modification.
In this paper, we neglect the problem concerning 
the surface integral term
(see Ref.~\refcite{gmm} for the serious treatment).
When we neglect the kinematic term and the potential term 
of $\vartheta$ in Eq.~(\ref{eq:lagrangian}), the action 
reduces to that of the Jackiw-Pi model.\cite{jp}
 From the variations in the action with respect to 
the metric $g_{\mu\nu}$ and the scalar field $\vartheta$, 
we obtain the field equations, respectively, 
\begin{eqnarray}
\label{eq:feq1}
 G^{\mu\nu} + \ell C^{\mu\nu} 
  & = & - 8\pi \left( T_{\rm m}^{\ \ \mu\nu} 
    + T_{\vartheta}^{\ \mu\nu} \right), \\
\label{eq:feq2}
 g^{\mu\nu} \nabla_{\mu} \nabla_{\nu} \vartheta
  & = & \frac{dV \left( \vartheta \right)}{d\vartheta}
   - \frac{\ell}{64\pi} \: ^{\ast} R^{\tau \ \mu\nu}_{\ \sigma}
   R^{\sigma}_{\ \tau\mu\nu} ,
\end{eqnarray}
where $G^{\mu\nu}$ is the Einstein tensor, 
$C^{\mu\nu}$ is the Cotton tensor defined by
\begin{eqnarray}
 C^{\mu\nu} & \equiv & - \frac{1}{2}
 \left[ \left( \nabla_{\sigma} \vartheta \right)
   \left( \varepsilon^{\sigma\mu\alpha\beta}
   \nabla_{\alpha} R^{\nu}_{\ \beta} + 
   \varepsilon^{\sigma\nu\alpha\beta}
   \nabla_{\alpha} R^{\mu}_{\ \beta} \right) +
   \left( \nabla_{\sigma} \nabla_{\tau} \vartheta \right)
   \left( \:\! ^{\ast} \! R^{\tau\mu\sigma\nu} 
    + \:\! ^{\ast} \! R^{\tau\nu\sigma\mu} \right) \right] ,
   \nonumber \\
\end{eqnarray}
$T_{\rm m}^{\ \ \mu\nu}$ is 
the energy-momentum tensor for matter, and $T_{\vartheta}^{\ \mu\nu}$ 
is the energy-momentum tensor of the scalar field, 
\begin{equation}
\label{eq:emt-vartheta}
 T_{\vartheta}^{\ \mu\nu}
 = \left( \nabla^{\mu} \vartheta \right)
   \left( \nabla^{\nu} \vartheta \right)
   - g^{\mu\nu} \left[ \frac{1}{2} ( \nabla^{\lambda} 
   \vartheta )
   \left( \nabla_{\lambda} \vartheta \right) 
   + V \left( \vartheta \right)
   \right].
\end{equation}
Thus Eqs.~(\ref{eq:feq1}) and (\ref{eq:feq2}) are basic equations
in the CS modified gravity.

\subsection{Basic equations for fluid flow}

We obtain basic equations for fluid motion 
under the CS modified gravity from the covariant 
divergence of Eq.~(\ref{eq:feq1}),
\begin{equation}
\label{eq:conserv}
 \nabla_{\nu} T_{\rm m}^{\ \ \mu\nu}
 = - \frac{\ell}{8\pi} \nabla_{\nu} C^{\mu\nu}
   - \nabla_{\nu} T_{\vartheta}^{\ \mu\nu}
 \equiv \Theta^{\mu} ,
\end{equation}
where $\Theta^{\mu}$ is defined as force
exerted on the ordinary matter. 
We can regard $\Theta^{\mu}$ as a CS correction.
Using Eq.~(\ref{eq:emt-vartheta}) and the equality\cite{jp} 
\begin{equation}
\label{eq:equality}
 \nabla_{\nu} C^{\mu\nu} 
  = \frac{1}{8} \left( \nabla^{\mu} \vartheta \right) \;\! 
    ^{\ast} \!R^{\sigma\ \nu\lambda}_{\ \tau} 
    R^{\tau}_{\ \sigma\nu\lambda} , 
\end{equation}
we obtain
\begin{eqnarray}
 \Theta_{\mu} \left( \vartheta \right)
 & = & \left( \nabla^{\nu} \vartheta \right)
       \left( \nabla_{\mu} \nabla_{\nu} 
       - \nabla_{\nu} \nabla_{\mu} \right) \vartheta .
\end{eqnarray}
Thus we find that $\Theta_{\mu}$ vanishes when the scalar 
field is regular everywhere. Then we derive
the usual equation of motion for matter, 
$\nabla_{\nu} T_{\rm m}^{\ \ \mu\nu} =0$. In this case, 
the motion of fluid is affected only through 
the change of the metric.
When we consider a test particle, 
we derive the usual geodesic equation from 
$\nabla_{\nu} T_{\rm m}^{\ \ \mu\nu}=0$, which is
independent of the mass. This fact means that 
the equivalence principle is valid for regular $\vartheta$
under the CS modified gravity. On the other hand, 
if there is a singularity in $\vartheta$, 
$\Theta_{\mu}$ would not vanish at the singularity. 
For example, if $\vartheta = \arctan (y/x) \equiv \phi $ 
in rectangular Cartesian coordinates, we derive 
\begin{equation}
 \Theta_{\mu} 
  = \left( 0, \frac{2\pi x}{x^2+y^2} \delta (x) \delta (y) ,
    \frac{2\pi y}{x^2+y^2} \delta (x) \delta (y) , 0 \right) ,
\end{equation}
where we have used the formula\cite{jp2}
\begin{equation}
 \left( \partial_{x} \partial_{y} 
       - \partial_{y} \partial_{x} \right) \phi
 = 2\pi \delta (x) \delta (y) .
\end{equation}
Here, $\delta (x)$ denotes the Dirac's delta function.
Therefore, when such a topological singularity exists, 
the force term $\Theta^{\mu}$ plays an important role
at the singularity.
Hereafter, we take account of the effect of 
the CS correction $\Theta^{\mu}$ on fluid motion.
We now assume perfect fluid again.
Equation (\ref{eq:conserv}) can be divided into two parts, 
i.e., the component parallel to $u^{\mu}$ and 
the components orthogonal to $u^{\mu}$.
The former gives the balance equation, i.e., 
the continuity equation with a source term,
\begin{equation}
\label{eq:g-c-e0}
 \left( \rho u^{\mu} \right)_{; \mu}
 + p u^{\mu}_{\ ;\mu} = - u_{\mu} \Theta^{\mu},
\end{equation}
where a semicolon denotes the covariant derivative.
The latter gives the equation of motion 
\begin{equation}
\label{eq:g-eom0}
 \left( \rho + p \right) u^{\nu} u^{\mu}_{\ ; \nu}
 + P^{\mu\nu} p_{,\nu} = P^{\mu}_{\ \ \nu} \Theta^{\nu} ,
\end{equation}
where $P^{\mu\nu}$ is the projection tensor 
onto the hypersurface orthogonal to $u^{\mu}$. 
Using the variables $s$ and $h$, we can write Eqs.~(\ref{eq:g-c-e0}) 
and (\ref{eq:g-eom0}), respectively, as 
\begin{equation}
\label{eq:csg-c-e}
 \left( s u^{\mu} \right)_{; \mu} = - {\cal S} ,
\end{equation}
\begin{equation}
\label{eq:csg-eom}
 u^{\nu} u^{\mu}_{\ \; ; \nu} + P^{\mu\nu} \left( \ln h\right)_{,\nu}
 = {\cal V}^{\mu} ,
\end{equation}
where the CS corrections ${\cal S}$ and ${\cal V}^{\mu}$ 
are defined by
\begin{eqnarray}
{\cal S} & \equiv & \frac{s}{\rho + p}
  u_{\mu} \Theta^{\mu} , \\
 {\cal V}^{\mu} & \equiv & \frac{1}{\rho + p}
  P^{\mu}_{\ \ \nu} \Theta^{\nu} .
\end{eqnarray}
The balance equation (\ref{eq:csg-c-e}) 
and the equation of motion (\ref{eq:csg-eom}) 
govern fluid dynamics under the CS modified gravity.

 From Eq.~(\ref{eq:csg-eom}), we also obtain the differential 
equation for vorticity $\omega^{\mu}$ as 
\begin{eqnarray}
\label{eq:csg-v-e}
 \left( \frac{h\omega^{\mu}}{s} \right)_{; \nu}
  \frac{u^{\nu}}{h} - \frac{h\omega^{\nu}}{s} 
  \left( \frac{u^{\mu}}{h}\right)_{; \nu} 
 = \frac{1}{s^2}{\cal S} \omega^{\mu}
  +  \frac{1}{s} \left[ {\cal V}_{\nu} \omega^{\nu} u^{\mu} 
  + \: ^{\ast} {\cal W}^{\mu\nu} u_{\nu} \right] , 
\end{eqnarray}
where 
\begin{equation}
 ^{\ast} {\cal W}^{\mu\nu} \equiv \frac{1}{2} 
   \varepsilon^{\mu\nu\lambda\sigma} {\cal V}_{\lambda ; \sigma} .
\end{equation}
Here $^{\ast} {\cal W}^{\mu\nu}$ is regarded as the rotational 
component of the force ${\cal V}^{\mu}$.
Thus, Eq.~(\ref{eq:csg-v-e}) governs the time evolution of 
vorticity in fluid motion under the CS modified gravity.

\subsection{Lagrangian description of fluid flow}

We discuss fluid motion from the viewpoint of 
the Lagrangian description.

We deal with the balance equation (\ref{eq:csg-c-e}). 
Applying the Lagrangian condition (\ref{eq:L-c}) to
Eq.~(\ref{eq:csg-c-e}), we derive
\begin{equation}
 \left( \sqrt{-g} s\right)_{, 0} = - \sqrt{-g}{\cal S}  .
\end{equation}
Thus we obtain 
\begin{equation}
\label{eq:csg-s}
 s\left( \tau , \mbox{\boldmath $x$} \right) 
 = \sqrt{\frac{g\left( \tau_{0} , \mbox{\boldmath $x$} \right)}
   {g\left( \tau , \mbox{\boldmath $x$} \right)}}  
   s\left( \tau_{0} , \mbox{\boldmath $x$} \right) 
   - \frac{1}{\sqrt{-g \left( \tau , \mbox{\boldmath $x$} \right)}} 
   \int_{\tau_{0}}^{\tau} d\tau' 
   \sqrt{-g}{\cal S} .
\end{equation}
When the pressure vanishes, Eq.~(\ref{eq:csg-s}) reduces to 
\begin{equation}
\label{eq:csg-rho}
 \rho \left( \tau , \mbox{\boldmath $x$} \right) 
 = \sqrt{\frac{g\left( \tau_{0} , \mbox{\boldmath $x$} \right)}
   {g\left( \tau , \mbox{\boldmath $x$} \right)}}  
   \rho \left( \tau_{0} , \mbox{\boldmath $x$} \right) 
   - \frac{1}{\sqrt{-g \left( \tau , \mbox{\boldmath $x$} \right)}} 
   \int_{\tau_{0}}^{\tau} d\tau' 
   \sqrt{-g}{\cal S} .
\end{equation}
 From Eq.~(\ref{eq:csg-s}) and (\ref{eq:csg-rho}), we see that 
if the CS Correction ${\cal S}$ vanishes, 
the entropy density $\sqrt{-g} s$ or the energy density 
$\sqrt{-g} \rho$ is conserved in the Lagrangian coordinates.
However, if the CS correction ${\cal S}$ does not vanish, 
$\sqrt{-g} s$ and $\sqrt{-g} \rho$ 
are no longer conserved. The function ${\cal S}$ 
provides a source of entropy or energy as seen in 
Eqs.~(\ref{eq:csg-s}) and (\ref{eq:csg-rho}).
Thus in general, the creation of entropy or energy may occur 
when there is a topological singularity in the scalar field
appearing in the CS modified gravity.

Next we discuss the vorticity equation (\ref{eq:csg-v-e})
using the Lagrangian condition (\ref{eq:L-c}).
As mentioned above, the temporal component of $\omega^{\mu}$
is not independent of the spatial components
because $\omega^{0} = g_{0i}\omega^{i}$. Hence we 
focus on the spatial components of $\omega^{\mu}$ again.
The spatial components of Eq.~(\ref{eq:csg-v-e}) give
\begin{equation}
  \left( \frac{h\omega^{i}}{s} \right)_{, 0}
  = \frac{h}{s^2}{\cal S} \omega^{i}
    + \frac{h}{s}  \: ^{\ast} {\cal W}^{i\nu}u_{\nu} , 
\end{equation}
Then we obtain
\begin{eqnarray}
\label{eq:csg-h-omega}
 h(\tau , \mbox{\boldmath $x$}) \omega^{i}(\tau , \mbox{\boldmath $x$})
 & = & 
   \frac{s(\tau,\mbox{\boldmath $x$})}{s(\tau_{0} ,\mbox{\boldmath $x$})} 
   h(\tau_{0} , \mbox{\boldmath $x$})
   \omega^{i}(\tau_{0} , \mbox{\boldmath $x$}) 
   + s(\tau,\mbox{\boldmath $x$}) \int_{\tau_{0}}^{\tau} d\tau' 
   \left( \frac{h}{s^2}{\cal S} \omega^{i} 
   + \frac{h}{s} \: ^{\ast} {\cal W}^{i\nu}u_{\nu} \right) .
 \nonumber \\
\end{eqnarray}
When the pressure vanishes, Eq.~(\ref{eq:csg-h-omega}) reduces to
\begin{eqnarray}
\label{eq:csg-omega}
 \omega^{i}(\tau , \mbox{\boldmath $x$})
 & = & 
  \frac{\rho(\tau,\mbox{\boldmath $x$})}
  {\rho (\tau_{0} ,\mbox{\boldmath $x$})}
   \omega^{i}(\tau_{0} , \mbox{\boldmath $x$})
  + \rho(\tau,\mbox{\boldmath $x$}) \int_{\tau_{0}}^{\tau} d\tau' 
  \left( \frac{1}{\rho^2}{\cal S} \omega^{i} 
  + \frac{1}{\rho} \: ^{\ast} {\cal W}^{i\nu}u_{\nu} \right) .
\end{eqnarray}
 From Eqs.~(\ref{eq:csg-h-omega}) and (\ref{eq:csg-omega}),
we see that if the CS corrections
${\cal S}$ and $^{\ast}{\cal W}^{i0}$ vanish, 
$h\omega^{i}/s$ or $\omega^{i}/\rho$ is conserved
in the Lagrangian coordinates.
However, if ${\cal S}\neq 0$ or $^{\ast}{\cal W}^{i\nu}\neq 0$, 
$h\omega^{i}/s$ or $\omega^{i}/\rho$ is no longer 
conserved. In particular, the second terms of the integrands 
in Eqs.~(\ref{eq:csg-h-omega}) and (\ref{eq:csg-omega}) may 
produce vorticity, regardless of the values of vorticity vector.
Therefore, even if the vorticity vanishes everywhere 
at the initial time, circulation may be created 
when there is a topological singularity in the scalar field
of the CS modified gravity.

\section{Summary}
\label{sec:summary}

We have considered fluid flow under the CS modified Maxwell theory
and under the CS modified gravity theory.
We investigated the effects of the CS corrections
on conserved quantities of relativistic fluid.
First of all, we obtained the CS corrections to the equations for fluid, 
i.e., the continuity equation, the equation of motion 
and the vorticity equation. For the discussion of vorticity, 
we introduced the four-vorticity in Eq.~(\ref{eq:4-v}). 
We also pointed out that the four-vorticity provides 
the expressions of both three-vorticity and fluid helicity 
density in the nonrelativistic limit. This fact means that
the four-vorticity unifies the physical pictures of 
circulation and fluid helicity in a relativistic framework.
To discuss conserved quantities of fluid, 
we utilized the Lagrangian description of fluid flow. 
In the CS modified Maxwell theory, we found that 
while the entropy or energy is conserved, 
the circulation and fluid helicity are not generally conserved.
We also found that the conservation of the magnetic helicity 
holds even in the CS modified Maxwell theory, while the 
conservation of the cross helicity does not hold in general.
In the CS modified gravity, when the CS scalar field
is regular everywhere, the CS corrections do not appear in the 
equations for fluid flow. In this case, we have 
the validity of the equivalence principle. 
This fact means that for a regular CS scalar field,
we cannot find the CS correction
directly from the test of equivalence principle.
Furthermore, we found that when there is a topological 
singularity in the CS scalar field, not only the circulation 
and fluid helicity but also the entropy and energy 
are not conserved in general at the singularity.

\section*{Acknowledgments}

We thank Prof.~Y.~Fukumoto for useful conversations. 
One author (K.K.) thanks Prof.~H.~Asada for fruitful discussions. 
Analytical calculations were performed in part 
on computers at YITP in Kyoto University.

\end{document}